\documentclass [12pt]{article}
\usepackage{graphicx,amssymb,amsfonts,latexsym,amsmath,amsthm,times}
\usepackage{subcaption}
\usepackage{epsfig}
\usepackage{fancyhdr}
\usepackage{color}
\usepackage[english]{babel}
\usepackage[style=numeric-comp, firstinits=true,backend=bibtex]{biblatex}
\renewbibmacro{in:}{%
  \ifentrytype{article}{}{\printtext{\bibstring{in}\intitlepunct}}}
\renewcommand{\theequation}{\thesection.\arabic{equation}}
\numberwithin{equation}{section}

       \renewcommand\theequation{\thesection\arabic{equation}}
\begin{document}


\thispagestyle{plain}

\vspace*{2cm} \normalsize \centerline{\Large \bf Proliferating L\'{e}vy Walkers and Front Propagation}

\vspace*{1cm}

\centerline{\bf H. Stage $^a$, S. Fedotov $^a$\footnote{Corresponding
author. E-mail: sergei.fedotov@manchester.ac.uk}
 and V. M\'{e}ndez$^b$}

\vspace*{0.5cm}

\centerline{$^a$School of Mathematics, The University of Manchester, M13 9PL Manchester, UK}

\centerline{$^b$Universitat Aut\'{o}noma de Barcelona, Departament de F\'{i}sica, Facultat de Ci\`{e}ncies. Edifici Cc}\centerline{08193-Cerdanyola del Vall\`{e}s (Barcelona) Spain}


\vspace*{1cm}

\noindent {\bf Abstract.}

We develop non-linear integro-differential kinetic equations for proliferating L\'{e}vy walkers with birth and death processes. A hyperbolic scaling is applied directly to the general equations to get the Hamilton-Jacobi equations that will allow to determine the rate of front propagation. We found the conditions for switching, birth and death rates under which the propagation velocity reaches the maximum value, i.e. the walker's speed. In the strong anomalous case the death rate was found to influence the velocity of propagation to fall below the walker's maximum speed.

\vspace*{0.5cm}


\noindent {\bf AMS subject classification:}


\vspace*{1cm}

\setcounter{equation}{0}
\section{Introduction}


L\'{e}vy walks have become of increasing interest in the past years, both mathematically and in the context of their many applications \cite{levyreview, sharks, bees, humans, qdots}. They are typical in describing anomalous superdiffusion and other transport phenomena.
A variety of experimental results indicate superdiffusive motion which is consistent with L\'{e}vy walks, including the spreading of cancer cells \cite{supercancer2011, supercancer2013}. L\'{e}vy walks with appropriate proliferation kinetics could therefore be used to model cancer propagation. So far modelling of L\'{e}vy walks has mainly been concerned with passive movements, and not proliferation. Some attempts have been made taking into account models with finite and random velocities \cite{velocities1, velocities2}. Alternative models have also been developed to describe biological or ecological movement, often in the form of composite continuous time random walks \cite{compmodel}.

The purpose of our paper is to develop a theory of proliferating L\'{e}vy walkers with birth and death kinetics, starting from a mesoscopic level. A L\'{e}vy walker has non-Markovian motion, which poses a challenge as we do not already have a non-linear integro-differential equation for proliferating walkers.
We are further interested in the front propagation in such systems. Our main technique is hyperbolic scaling applied from the onset without long-time large scale approximation. This is a known method previously applied to problems of front propagation involving non-Markovian transport \cite{SergeiBook}. Our reasoning for this is as follows.
The standard method to treat kinetic equations is to apply parabolic scaling and obtain a diffusion equation with diffusion coefficient  $D=v^2T$ where $v$ is the velocity and $T$ mean running time \cite{OthmerHillen, HillenDL2000}. If we na\"{i}vely take into account birth and death processes, we obtain the well-known Fisher equation \cite{SergeiBook, murray}
\begin{equation}
\label{eq:eqn-f}
\frac{\partial\rho}{\partial t}=D\frac{\partial^2\rho}{\partial x^2}+r\rho(1-\rho),
\end{equation}
where the propagation speed of the travelling wave emerging from an initial condition with compact support is 
\begin{equation}
\label{eq:standard-v}
u=2\sqrt{Dr}=2v\sqrt{Tr}.
\end{equation}
It is clear that at $rT=1$, the propagation speed is $2v$, which contradicts the maximum possible speed $v$ of the walker and violates the causality principle.
This difficulty can be resolved using hyperbolic corrections to \eqref{eq:eqn-f}. The corresponding equation of motion \cite{SergeiBook} (p.38) is
\begin{equation}
T\frac{\partial^2\rho}{\partial t^2}+\left[1-T F'(\rho)\right]\frac{\partial\rho}{\partial t}=D\frac{\partial^2\rho}{\partial x^2}+F(\rho),
\end{equation}
which is the reaction-telegraph equation with non-linear kinetic term $F(\rho)=r\rho(1-\rho)$. This has propagation speed
\begin{equation}
\label{eq: fischer}
u=\frac{\sqrt{4Dr}}{1+rT}=\frac{2v\sqrt{rT}}{1+rT},
\end{equation}
for $rT<1$ and $u=v$ when $rT=1$ \cite{fort, SergeiBook}. 
For this reason, our intention in this paper is to study a non-Markovian kinetic equation for proliferating L\'{e}vy walkers without applying diffusion approximations. Markovian kinetic equations have been considered in \cite{bouin, bouin1}.

As we do not have a general equation for the density of proliferating walkers with arbitrary running time probability distribution, we cannot simply introduce a modification in the form of an additional reaction term as in \eqref{eq:eqn-f}. Similar problems have been solved recently for subdiffusive proliferating walkers \cite{subdiffreview, sergeisubdiff2010, sergeisubdiff2015, strakesubdiff2015,yadav}. In this paper we focus on superdiffusive kinetic models. The challenge is to implement the birth-death kinetic terms into non-Markovian L\'{e}vy models. Our following main objective is finding the equation for the density of proliferating L\'{e}vy walkers. Our specific goals are then to use this equation to find the resulting travelling wave and rate of front propagation. It should then be possible to find conditions under which the propagation velocity is upper bounded.
We intend to show that this mesoscopic outset of the model does not necessarily lead to the same macroscopic velocity as if we were to use standard diffusion methods.
 
Our method has the benefit of applying to various forms of the running time probability distribution, which is of interest given recent evidence that this may not be exponential \cite{levybacteria}. Our definition of L\'{e}vy walks consequently considers distributions that are not exponential, thus avoiding the reduction to Markovian form which otherwise occurs. We make a distinction between L\'{e}vy flights where individuals make the displacement instantaneously and rest in that position for some time, and walks where the displacement is continuous. The former again reduces the problem to be Markovian.

\vspace*{0.5cm}
\setcounter{equation}{0}
\section{Proliferating L\'{e}vy Walkers}

For simplicity we will treat one-dimensional L\'{e}vy walks with walkers moving in the $x$-direction with constant velocity $v$. We assume that walkers change their direction with rate $\gamma(\tau)$ depending on the running time $\tau$:
\begin{equation}
\gamma(\tau)=\frac{\psi(\tau)}{\Psi(\tau)},
\label{eq: gamma-def}
\end{equation}
where $\psi(\tau)$ is the probability density function (PDF) of running times and $\Psi(\tau)$ the corresponding survival probability of the walker $\int_t^\infty\psi(\tau)d\tau$ \cite{sergei2015ballistic}. In this paper we consider L\'{e}vy walkers to have an arbitrary running time distribution. In Section 5 we particularly examine distributions with divergent moments. To model persistent movement, we assume that $\gamma(\tau)$ is a decreasing function with running time. Hence, the longer a walker has been moving in a certain direction, the less likely it is to change.
For constant $\gamma(\tau)$, we obtain an exponential distribution of running times \cite{expruntime}.\newline
Our intention is to derive a kinetic equation for the mean total density
\begin{equation}
\label{eq: totalrho}
\rho(x,t)=\rho_+(x,t)+\rho_-(x,t),
\end{equation}
where $\rho_\pm(x,t)$ are the densities of individuals moving to the right (+) and left (-).\newline
Our challenge now is to implement the birth and death rates into this non-Markovian transport process. We assume that the production or birth term for walkers moving to the right is
\begin{equation}
f^+(\rho)(\kappa\rho_++(1-\kappa)\rho_-),
\end{equation}
and to the left
\begin{equation}
f^+(\rho)(\kappa\rho_-+(1-\kappa)\rho_+).
\end{equation}
The equations account for correlations in the direction of motion of the 'mother' individual and corresponding 'daughter' individuals. The degree of correlation, $\kappa$, takes three notable values. At $\kappa=1/2$, there is no correlation. At $\kappa=1$ we have complete correlation whereas $\kappa=0$ corresponds to complete anticorrelation \cite{SergeiBook} (p.43).
The total birth term for the mean density is thus expressed as
\begin{equation}
f^+(\rho)(\kappa\rho+(1-\kappa)\rho).
\end{equation}
The corresponding death terms for the walkers are
\begin{equation}
\theta\rho_\pm,
\end{equation}
where $\theta>0$ is the death rate. The net production of walkers is
\begin{equation}
(f^+(\rho)-\theta)\rho.
\end{equation}
In order to sustain the travelling wave solution, we assume
\begin{equation}
f^+(\rho)-\theta>0. 
\label{eq: growth}
\end{equation}
We further assume that the net production rate takes the maximum value at $\rho=0$, and takes zero value at $\rho=1$ so the mean total density varies as follows $0\leq\rho(x,t)\leq 1$, where $\rho=1$ corresponds to saturation of the system.

In what follows we consider only the case of complete correlation and no correlation. 
We are concerned with birth terms which contain one stable $\rho=1$ and one unstable $\rho=0$ solution. A well-known example of such is logistic birth of the form $\rho(1-\rho)$, though the expression may be expanded to contain higher orders in $\rho$. At saturation, the net production of walkers must be zero so that we find, setting $\rho=1$,
\begin{equation}
(f^+(1)-\theta)=0,
\end{equation}
regardless of the value of $\kappa$. In the (low density) unstable state, the birth term has no such restrictions and is free to grow at a constant rate
\begin{equation}
f^+(0)=\text{constant}.
\label{eq: free-growth}
\end{equation}
In the case of sustained growth described by \eqref{eq: growth} the density must be below saturation. Travelling waves will consequently appear, travelling from the stable into the unstable state (see Figure \ref{fig: scale}). In what follows we consider three different models for front propagation into the unstable state. 

\subsection{Models A, B and C}
In this subsection we consider three models corresponding to different assumptions regarding the birth correlations and ageing. The reason for this is the evidence we have regarding the importance of such assumptions in previously examined non-Markovian models for subdiffusive motion \cite{subdiffreview, sergeisubdiff2010, sergeisubdiff2015, strakesubdiff2015,yadav}.\newline
\textbf{Model A}

We first consider the case $\kappa=1$, corresponding to complete correlation such that (newborn) daughter individuals move in the same direction as the (producing) mother individual. We assume the death rate $\theta$ is constant and the birth rate $f^+(\rho)$ is dependent on the mean total density. This can describe both competition in highly dense areas or freely available resources in less occupied areas should such systems be of interest. The governing equation for the densities $\rho_\pm$ can be written as
\label{sec:easy}
\begin{equation}
\label{eq:rho1-b}
\frac{\partial\rho_+}{\partial t}+v\frac{\partial \rho_+}{\partial x}=-\frac{1}{2}(i_+ - i_-) -\theta\rho_+ + f^+(\rho)\rho_+,
\end{equation}
\begin{equation}
\label{eq:rho2-b}
\frac{\partial\rho_-}{\partial t}-v\frac{\partial \rho_-}{\partial x}=\frac{1}{2}(i_+ - i_-) -\theta\rho_- + f^+(\rho)\rho_-,
\end{equation}
where we note the birth terms only contribute to the respective direction of movement. The interaction terms $ i_\pm$ are defined below in \eqref{eq:interaction}, and can be understood physically as the flux of walkers occurring when they they have finished walking in their currently specified direction. That is, the walkers do not constantly change direction as they have specified running times, and so the flux only occurs when the current run is completed. See Appendix A for the details of the derivation. Furthermore, the newborn walkers are initialised with a running time of $\tau=0$.\newline 
The interaction/exchange terms accounting for the changes in direction \cite{sergei2015ballistic}
\begin{equation}
\label{eq:interaction}
i_{\pm}(x,t)=\int_0^tK(t-\tau)\rho_{\pm}(x\mp v(t-\tau),\tau)e^{-\theta(t-\tau)}d\tau
\end{equation}
depend on the density of walkers moving in a particular direction for all running times between the beginning $t=0$ and $t=\tau$, i.e. all walkers which have travelled from other positions for any amount of time to now be  at $(x,t)$. This clearly demonstrates the non-Markovian nature of the model. As the walkers may die at any point throughout their walk the term in $\theta$ arises naturally, but $f^+$ does not appear as walkers are produced with zero running time. Finally, the exchange terms also depend on the PDF of running times $\psi(\tau)$ via the kernel $K$ in Laplace space:
\begin{equation}
\label{eq: k}
\widehat{K}(s)=\frac{\widehat{\psi}(s)}{\widehat{\Psi}(s)}=\frac{s\widehat{\psi}(s)}{1-\widehat{\psi}(s)}.
\end{equation}

It is important to note that \eqref{eq:interaction} was obtained by integration to remove the dependency of the equations on $\tau$. This involves convolutions in Fourier-Laplace space via the transformation
\begin{equation}
\mathcal{FL}[f(x,t)]=\tilde{f}(k,s)=\int_\mathbb{R}\int_0^\infty e^{ikx-st}f(x,t)dtdx.
\end{equation}
Consequently, seemingly small changes to the production terms in the equations of motion may lead to structurally different solutions for $\rho(x,t)$. To clarify the ideas introduced here, we outline simple examples with specific kernels in Section \ref{sec:examples}\newline\newline
\textbf{Model B}

In contrast to our previous model, we now consider the uncorrelated case $\kappa=1/2$ where the direction of movement mother and daughter walkers is not correlated at all. In this case, half of the produced individuals are sent in each direction (still with zero running time). This changes the equations for the densities to
\begin{equation}
\frac{\partial\rho_+}{\partial t}+v\frac{\partial \rho_+}{\partial x}=-\frac{1}{2}(i_+ - i_-) -\theta\rho_+ + f^+(\rho)\frac{\rho_++\rho_-}{2},
\end{equation}
\begin{equation}
\frac{\partial\rho_-}{\partial t}-v\frac{\partial \rho_-}{\partial x}=\frac{1}{2}(i_+ - i_-) -\theta\rho_- + f^+(\rho)\frac{\rho_++\rho_-}{2},
\end{equation}
and an exchange term as in \eqref{eq:interaction}. The derivation is similar to that of Appendix A. Although these are deceptively similar to \eqref{eq:rho1-b}, \eqref{eq:rho2-b} the equations are now further coupled. The consequences of this are explored later in this paper.\newline\newline
\textbf{Model C}

Finally, we consider the case of a constant birth rate as described by \eqref{eq: free-growth} with a running time equal to that of the walker which produced it in addition to $\kappa=1$. This means that there is a correlation not only between the direction of movement of the mother and daughter walkers, but also their running times. For convenience we write this constant as $f^+(0)=f^+$. The governing equations are now
\begin{equation}
\frac{\partial\rho_+}{\partial t}+v\frac{\partial \rho_+}{\partial x}=-\frac{1}{2}(i_+ - i_-) -\theta\rho_+ + f^+\rho_+,
\end{equation}
\begin{equation}
\frac{\partial\rho_-}{\partial t}-v\frac{\partial \rho_-}{\partial x}=\frac{1}{2}(i_+ - i_-) -\theta\rho_- + f^+\rho_-,
\end{equation}
where the derivation is again similar to that of Appendix A, but the interaction terms are necessarily different. For this model, is is possible to simplify the rates to a net growth rate for the system $\delta=f^+-\theta$. The interaction term becomes
\begin{equation}
\label{eg: interaction2}
i_{\pm}(x,t)=\int_0^tK(t-\tau)\rho_{\pm}(x\mp v(t-\tau),\tau)e^{(f^+-\theta)(t-\tau)}d\tau.
\end{equation}
Clearly in this case there is a correlation between the transport and production, which does not appear in the previously introduced models. Having introduced the details of our three models, we now proceed to illustrate the effect of memory kernels on Model A. 

\subsection{Examples of Model A}
\label{sec:examples}
In this subsection we are going to consider two examples, one Markovian and one Non-Markovian. The purpose of this is to illustrates the coupling resulting from a non-exponential memory kernel.\newline
\textbf{Markovian case}\newline
In the particular case when the switching rate $\gamma$ is constant, $\psi(\tau)=\gamma e^{-\gamma\tau}$ the distribution of running times is exponential. We then recover the Markovian model
\begin{equation}
\label{eq: markov1}
\frac{\partial\rho_+}{\partial t}+v\frac{\partial \rho_+}{\partial x}=-\frac{\gamma}{2}(\rho_+ - \rho_-) + (f^+-\theta)\rho_+,
\end{equation}
\begin{equation}
\label{eq: markov2}
\frac{\partial\rho_-}{\partial t}-v\frac{\partial \rho_-}{\partial x}=\frac{\gamma}{2}(\rho_+ - \rho_-) + (f^+-\theta)\rho_-.
\end{equation}
In this case, one can consider a net growth term $\delta=f^+-\theta>0$ without needing to worry about the subtleties in the running time dependencies. These correspond with a telegraph equation for $f^+,\theta=0$ and is well-documented \cite{telegraph1, telegraph2}.\newline\newline
\textbf{Non-Markovian case}\newline
If we instead of an exponential distribution consider $\psi(\tau)=\lambda^2t e^{-\lambda t}$ corresponding to a gamma-distribution, equations \eqref{eq: markov1}, \eqref{eq: markov2} become
\begin{equation}
\frac{\partial\rho_+}{\partial t}+v\frac{\partial \rho_+}{\partial x}=-\frac{1}{2}(i_+ - i_-) + (f^+-\theta)\rho_+,
\end{equation}
\begin{equation}
\frac{\partial\rho_-}{\partial t}-v\frac{\partial \rho_-}{\partial x}=\frac{1}{2}(i_+ - i_-) + (f^+-\theta)\rho_-,
\end{equation}
where the interaction terms (using \eqref{eq: k}) are now written as
\begin{equation}
i_{\pm}(x,t)=\lambda^2\int_0^te^{-2\lambda\tau}e^{-\theta\tau}\rho_{\pm}(x\mp v\tau,t-\tau)d\tau.
\end{equation}
This case is now clearly non-Markovian. Furthermore, it is trivially seen that the death rate couples with the memory kernel by virtue of both being expressed in exponential form, thus changing its effective form. Examples for power law distributed kernels will be illustrated later. 

\vspace*{0.5cm}
\setcounter{equation}{0}
\section{Single equation for the total density}
We now proceed to obtaining a single equation for the total density in each of the introduced models. Our method is similar to ones used in previous work \cite{newsergei}. Defining a flux term
\begin{equation}
J(x,t)=v\rho_+(x,t)-v\rho_-(x,t)
\end{equation}
of the directionality of movement and using \eqref{eq: totalrho}, we obtain
\begin{equation}
\rho_{\pm}(x,t)=\frac{1}{2}\left(\rho(x,t)\pm \frac{J(x,t)}{v}\right).
\label{eq: rhopm}
\end{equation}
In what follows our intention is to consider the propagation of the front into the unstable state. It is well known that in such cases we need only consider the linear terms in the birth rates as additional terms vanish under our intended hyperbolic scaling. This corresponds to the unstable $\rho=0$ state discussed in \eqref{eq: free-growth}. Consequently we can conveniently write
\begin{equation}
f^+(\rho)=r=\text{const.}
\end{equation}
This holds even if the original expressions for $f^+(\rho)$ are non-linear, so long as they obey the properties outlined in Section 2. 
\newline\newline
\textbf{Model A}

By expression of \eqref{eq:rho1-b}, \eqref{eq:rho2-b} in terms of $\rho, J$ and cross-differentiation to remove the flux, we get two different expressions for the differential equation (see Appendix B for details). If $f^{+'}(\rho)=0$, i.e. for a constant birth rate $r$, we get the expression
\begin{equation}
\label{eq:diffeq}
\frac{\partial^2\rho}{\partial t^2}-v^2\frac{\partial^2\rho}{\partial x^2}-v\frac{\partial}{\partial x}(i_+-i_-)+(r-\theta)^2\rho-2(r-\theta)\frac{\partial\rho}{\partial t}=0.
\end{equation}
A more involved expression can be derived for a general birth rate $f^+(\rho)$, which we neglect to include here as the subsequent scaling is known to eliminate the extra terms which would arise in \eqref{eq:diffeq}. Using \eqref{eq:interaction}, one obtains the final integro-differential equation for the constant birth rate
\begin{equation}
\label{eq:start}
\begin{split}
\frac{\partial^2\rho}{\partial t^2}-v^2\frac{\partial^2\rho}{\partial x^2}+2(\theta -r)\frac{\partial\rho}{\partial t}&+(\theta - r)^2\rho
=\\
&-\frac{1}{2}\int_0^tK(z)e^{-\theta z}\left(\frac{\partial}{\partial t}+\theta-r-v\frac{\partial}{\partial x}\right)\rho(x-vz,t-z)dz\\
&-\frac{1}{2}\int_0^tK(z)e^{-\theta z}\left(\frac{\partial}{\partial t}+\theta-r+v\frac{\partial}{\partial x}\right)\rho(x+vz,t-z)dz.
\end{split}
\end{equation}
This is one of our main results which will be used to find the equation for the front propagation of the system.\newline\newline
\textbf{Model B}

Using the same method as above, we obtain the non-linear equation
\begin{equation}
\frac{\partial^2\rho}{\partial t^2}-v^2\frac{\partial^2\rho}{\partial x^2}-v\frac{\partial}{\partial x}(i_+-i_-)-\theta(f^+(\rho)-\theta)\rho+(2\theta-f^+(\rho)-\rho f^{+'}(\rho))\frac{\partial\rho}{\partial t}=0.
\end{equation}
Using \eqref{eq:interaction}, this has full integro-differential form
\begin{equation}
\label{eq: full-b}
\begin{split}
\frac{\partial^2\rho}{\partial t^2}-v^2\frac{\partial^2\rho}{\partial x^2}+(2\theta-&f^+(\rho)-\rho f^{+'}(\rho))\frac{\partial\rho}{\partial t}-\theta(f^+(\rho)-\theta)\rho\\
=&-\frac{1}{2}\int_0^tK(z)e^{-\theta z}\left(\frac{\partial}{\partial t}+\theta-f^+(\rho)-v\frac{\partial}{\partial x}\right)\rho(x-vz,t-z)dz\\
&-\frac{1}{2}\int_0^tK(z)e^{-\theta z}\left(\frac{\partial}{\partial t}+\theta-f^+(\rho)+v\frac{\partial}{\partial x}\right)\rho(x+vz,t-z)dz.
\end{split}
\end{equation}
We note that depending on both the forms of $f^+(\rho)$ and $K(z)$, these equations may be highly non-linear. However, as discussed above, the influential terms will be constant, simplifying \eqref{eq: full-b}.\newline\newline
\textbf{Model C}

By a similar method to before where we assume a constant production rate $r$ from the onset,
\begin{equation}
\frac{\partial^2\rho}{\partial t^2}-v^2\frac{\partial^2\rho}{\partial x^2}-v\frac{\partial}{\partial x}(i_+-i_-)+(r-\theta)^2\rho-2(r-\theta)\frac{\partial\rho}{\partial t}=0,
\end{equation}
where by \eqref{eg: interaction2}, we obtain a final equation of the form
\begin{equation}
\label{eq: hj-c}
\begin{split}
\frac{\partial^2\rho}{\partial t^2}-v^2\frac{\partial^2\rho}{\partial x^2}+2(\theta -r)&\frac{\partial\rho}{\partial t}+(\theta - r)^2\rho
=\\
&-\frac{1}{2}\int_0^tK(z)e^{(r-\theta)z}\left(\frac{\partial}{\partial t}+\theta-r-v\frac{\partial}{\partial x}\right)\rho(x-vz,t-z)dz\\
&-\frac{1}{2}\int_0^tK(z)e^{(r-\theta)z}\left(\frac{\partial}{\partial t}+\theta-r+v\frac{\partial}{\partial x}\right)\rho(x+vz,t-z)dz.
\end{split}
\end{equation}

While the correlations are seemingly trivial changes to the equations of motion, they lead to structurally different solutions for the front propagation.

\vspace*{0.5cm}
\setcounter{equation}{0}
\section{Front Propagation}
We are now in a position to find the rate of the travelling wave solutions. To ensure a minimum propagation speed we will study the equations with a front-like initial condition
\begin{equation}
\rho_0(x)=\begin{cases}
1, & \text{if } x<0\\
0, & \text{if } x>0.
\end{cases}
\end{equation}
\begin{figure}
\captionsetup[subfigure]{width=0.9\textwidth}
\centering
\begin{subfigure}{0.45\textwidth}
\centering
\includegraphics[scale=0.5]{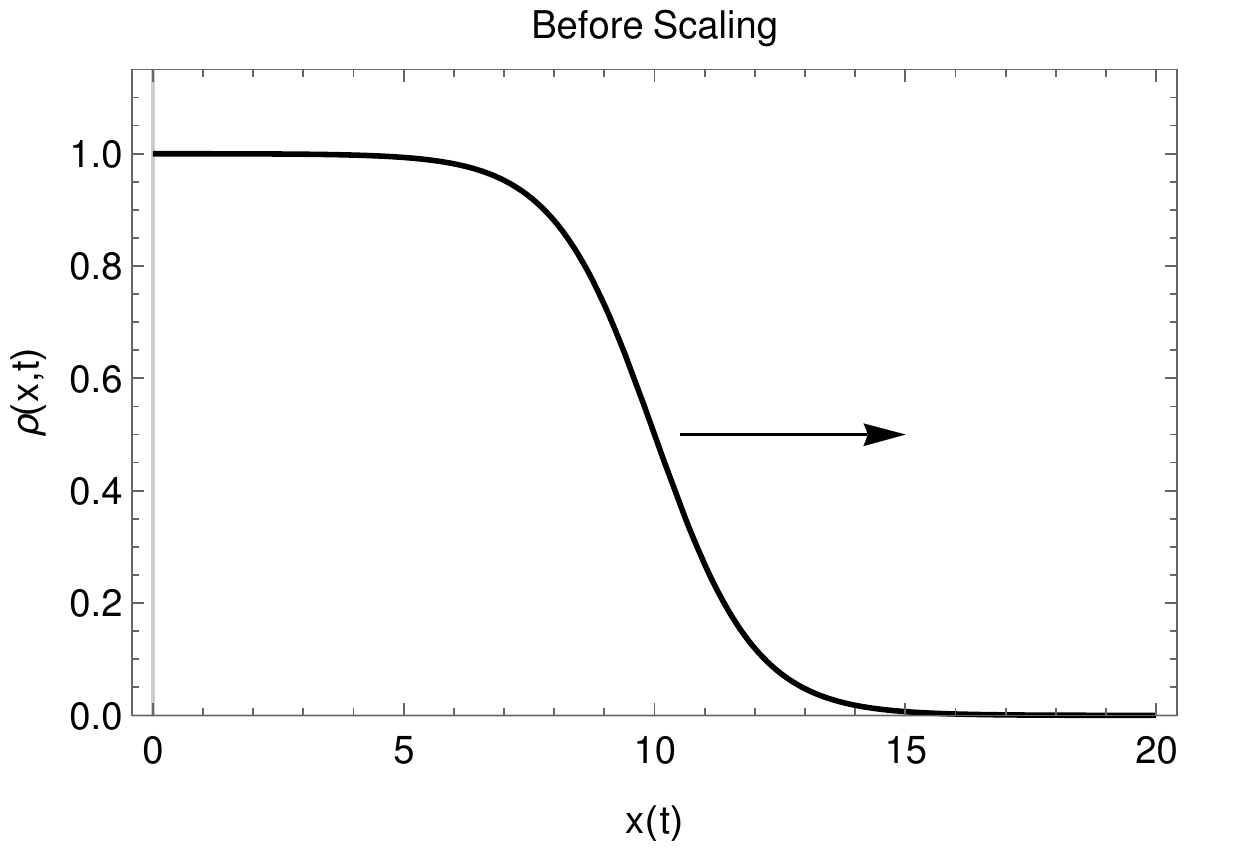}
\caption{The propagating front before hyperbolic scaling is applied. Smooth transition between $\rho=1$ and $\rho=0$.}
\end{subfigure}%
\begin{subfigure}{0.45\textwidth}
\centering
\includegraphics[scale=0.5]{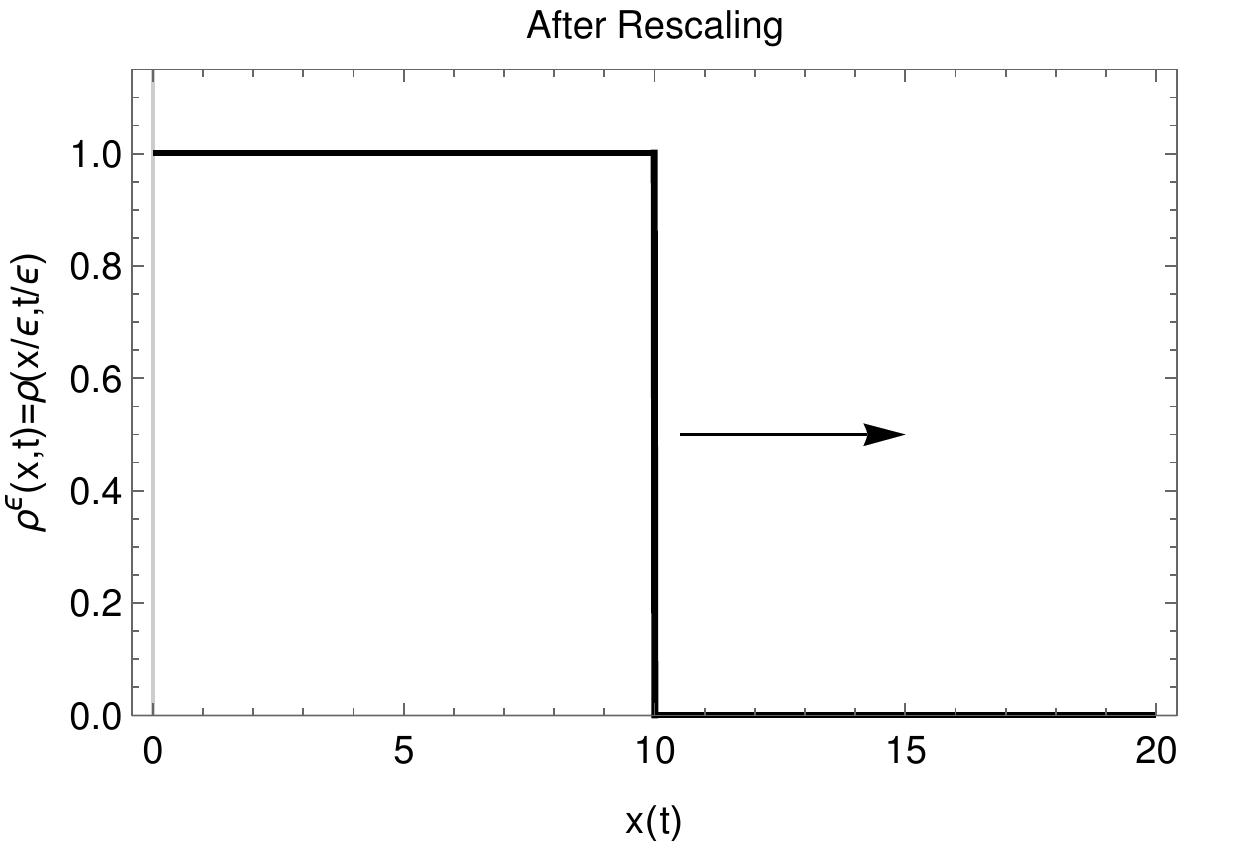}
\caption{The propagating front after hyperbolic scaling is applied. Step-like transition between $\rho=1$ and $\rho=0$.}
\label{fig: b}
\end{subfigure}
\caption{Graphical illustration of the effect of hyperbolic scaling on the propagation front.}
\label{fig: scale}
\end{figure}

For the models we are considering, the speed of propagation is independent of the profile. That is, the speed at which the front propagates does not explicitly depend on the front and thus remains unchanged under scaling. Consequently, we choose to apply a hyperbolic scaling in order to reduce the front to a geometric structure from which the velocity can easily be found \cite{SergeiBook} (p.155). We now proceed to apply a hyperbolic scaling $x\rightarrow \frac{x}{\epsilon},\quad t\rightarrow \frac{t}{\epsilon }$ and introduce a rescaled density $\rho ^{\epsilon}(x,t)=\rho \left(\frac{x}{\epsilon },\frac{t}{\epsilon }\right)$ which becomes the
reaction front in the limit $\epsilon \rightarrow 0$ (see Figure \ref{fig: b}). We then apply an exponential transformation to $\rho $ 
\begin{equation}
t\rightarrow \frac{t}{\epsilon },\qquad x\rightarrow \frac{x}{\epsilon}
,\qquad \rho ^{\epsilon }(x,t)=\exp \left( -\frac{G(x,t)}{\epsilon }\right) ,
\label{eq: scale}
\end{equation}
where the full form or any non-linearities in $\rho$ would be preserved for more
involved models.
We define the reaction front position $x(t)$ as the point in space at which our rescaled field $\rho^\epsilon$ changes from 1 to 0 (see Figure \ref{fig: scale}). It follows from \eqref{eq: scale} that the position for the reaction front $x(t)$ can be found from $G(x(t),t)=0$. 
It is clear that provided $\lim_{\epsilon\rightarrow 0}G(x,t)>0$ we obtain $\lim_{\epsilon\rightarrow 0}\rho ^{\epsilon }(x,t)=0$. Otherwise, when $G(x,t)=0,\lim_{\epsilon \rightarrow 0}\rho ^{\epsilon }(x,t)=1$. \cite{sergei2001frontvel, SergeiBook}\newline
Our reasoning for ignoring non-linear terms now becomes clear. Under the aforementioned limits as $\epsilon\rightarrow0$, the birth terms simplify to linear terms $r=f^+(0)$ as at the front we can set $\rho^\epsilon=0$ (giving $f^+(\rho^\epsilon)=f^+(0)$).\newline\newline
\textbf{Model A}

Applying hyperbolic scaling \eqref{eq: scale} to Model A \eqref{eq:start}, we obtain
\begin{equation}
\label{eq:fullhyperbolic}
\begin{split}
\epsilon^2\frac{\partial^2\rho^{\epsilon}}{\partial t^2}-\epsilon^2v^2\frac{\partial^2\rho^{\epsilon}}{\partial x^2}&+2(\theta -r)\epsilon\frac{\partial\rho^{\epsilon}}{\partial t}+(\theta - r)^2\rho^{\epsilon}=\\
&-\frac{1}{2}\int_0^{t/\epsilon}K(\tau)e^{-\theta\tau}\left(\epsilon\frac{\partial}{\partial t}+\theta-r-v\epsilon\frac{\partial}{\partial x}\right)\rho^{\epsilon}\left(\frac{x}{\epsilon}-v\tau,\frac{t}{\epsilon}-\tau\right)d\tau\\
&-\frac{1}{2}\int_0^{t/\epsilon}K(\tau)e^{-\theta\tau}\left(\epsilon\frac{\partial}{\partial t}+\theta-r+v\epsilon\frac{\partial}{\partial x}\right)\rho^{\epsilon}\left(\frac{x}{\epsilon}+v\tau,\frac{t}{\epsilon}-\tau\right)d\tau.
\end{split}
\end{equation}
The density terms in the integrals can be written
\begin{equation}
\rho^{\epsilon}\left(\frac{x}{\epsilon}\mp v\tau,\frac{t}{\epsilon}-\tau\right)=\exp\left(-\frac{G(x\mp\epsilon v\tau,t-\epsilon\tau)}{\epsilon}\right)\approx\exp\left(-\frac{G}{\epsilon}+\tau\frac{\partial G}{\partial t}\pm v\tau\frac{\partial G}{\partial x}\right),
\end{equation}
where the subsequent terms vanish in the limit $\epsilon\to0$. 
Substitution of \eqref{eq: scale} into \eqref{eq:fullhyperbolic} and expanding to first order in $\epsilon $, give an equation for $G(x,t)$
\begin{equation}
\begin{split}
\left(\frac{\partial G}{\partial t}\right)^2-v^2\left(\frac{\partial G}{\partial x}\right)^2-2&(\theta -r)\frac{\partial G}{\partial t}+(\theta - r)^2=\\
&-\frac{1}{2}\int_0^\infty K(\tau)e^{-\theta\tau}\left(-\frac{\partial G}{\partial t}+\theta-r+v\frac{\partial G}{\partial x}\right)e^{\tau\left(\frac{\partial G}{\partial t}+v\frac{\partial G}{\partial x}\right)}d\tau\\
&-\frac{1}{2}\int_0^\infty K(\tau)e^{-\theta\tau}\left(-\frac{\partial G}{\partial t}+\theta-r-v\frac{\partial G}{\partial x}\right)e^{\tau\left(\frac{\partial G}{\partial t}-v\frac{\partial G}{\partial x}\right)}d\tau.
\end{split}
\end{equation}
The advantage of this equation is that it can be expressed in terms of the Laplace transformation of the memory kernel. This naturally leads to the formulation
\begin{equation}
\label{eq:elimit-b}
\begin{split}
\left(\frac{\partial G}{\partial t}\right)^2-v^2\left(\frac{\partial G}{\partial x}\right)^2-2(\theta -r)&\frac{\partial G}{\partial t}+(\theta - r)^2=\\
&-\frac{1}{2}\left(-\frac{\partial G}{\partial t}+\theta-r+v\frac{\partial G}{\partial x}\right)\widehat{K}\left(\theta-\frac{\partial G}{\partial t}-v\frac{\partial G}{\partial x}\right)\\
&-\frac{1}{2}\left(-\frac{\partial G}{\partial t}+\theta-r-v\frac{\partial G}{\partial x}\right)\widehat{K}\left(\theta-\frac{\partial G}{\partial t}+v\frac{\partial G}{\partial x}\right).
\end{split}
\end{equation}
This is the generalised, relativistic Hamilton-Jacobi equation corresponding to \eqref{eq:start}. An alternative method for finding this equation is demonstrated in Appendix C. By solving this equation for $G(x,t)=0$, we will be able to find the front position for the model (see Figure \ref{fig: b}).\newline\newline
\textbf{Model B}

Applying the same method as before to \eqref{eq: full-b}, we obtain the expression
\begin{equation}
\begin{split}
\left(\frac{\partial G}{\partial t}\right)^2-v^2\left(\frac{\partial G}{\partial x}\right)^2-(2\theta -r)&\frac{\partial G}{\partial t}+\theta(\theta - r)=\\
&-\frac{1}{2}\left(-\frac{\partial G}{\partial t}+\theta-r+v\frac{\partial G}{\partial x}\right)\widehat{K}\left(\theta-\frac{\partial G}{\partial t}-v\frac{\partial G}{\partial x}\right)\\
&-\frac{1}{2}\left(-\frac{\partial G}{\partial t}+\theta-r-v\frac{\partial G}{\partial x}\right)\widehat{K}\left(\theta-\frac{\partial G}{\partial t}+v\frac{\partial G}{\partial x}\right).
\end{split}
\end{equation}
We note that in this model, the birth rate appears in a different way than observed in the previous case. This has different implications for the propagation rate (see subsection \ref{sec: kernel}).\newline\newline
\textbf{Model C}

We now use \eqref{eq: hj-c} to find the Hamilton-Jacobi equation for Model C.
\begin{equation}
\begin{split}
\left(\frac{\partial G}{\partial t}\right)^2-v^2\left(\frac{\partial G}{\partial x}\right)^2-2&(\theta -r)\frac{\partial G}{\partial t}+(\theta - r)^2=\\
&-\frac{1}{2}\left(-\frac{\partial G}{\partial t}+\theta-r+v\frac{\partial G}{\partial x}\right)\widehat{K}\left(\theta-r-\frac{\partial G}{\partial t}-v\frac{\partial G}{\partial x}\right)\\
&-\frac{1}{2}\left(-\frac{\partial G}{\partial t}+\theta-r-v\frac{\partial G}{\partial x}\right)\widehat{K}\left(\theta-r-\frac{\partial G}{\partial t}+v\frac{\partial G}{\partial x}\right).
\end{split}
\end{equation}
This can be expressed in terms of an effective growth rate $\delta=r-\theta$, as these two rates now always appear together.
\begin{equation}
\begin{split}
\left(\frac{\partial G}{\partial t}+\delta\right)^2-v^2\left(\frac{\partial G}{\partial x}\right)^2=&\frac{1}{2}\left(\frac{\partial G}{\partial t}+\delta-v\frac{\partial G}{\partial x}\right)\widehat{K}\left(-\delta-\frac{\partial G}{\partial t}-v\frac{\partial G}{\partial x}\right)\\
+&\frac{1}{2}\left(\frac{\partial G}{\partial t}+\delta+v\frac{\partial G}{\partial x}\right)\widehat{K}\left(-\delta-\frac{\partial G}{\partial t}+v\frac{\partial G}{\partial x}\right).
\end{split}
\end{equation}

\subsection{Hamiltonian and Generalised Momentum}
It is convenient to rewrite the Hamilton-Jacobi equations in terms of the Hamiltonian $H$ and generalised momentum $p$
\begin{equation}
\label{eq: defns}
H=-\frac{\partial G}{\partial t}, \qquad\qquad p=\frac{\partial G}{\partial x},
\end{equation}
which will be of further use later when determining the front velocity \cite{sergei2001frontvel}. Consequently, Model A \eqref{eq:elimit-b} becomes
\begin{equation}
\label{eq:laplaceform}
\begin{split}
\left(H+\theta-r\right)^2-v^2p^2=-&\frac{H+\theta-r+vp}{2}\widehat{K}(H+\theta-vp)-\frac{H+\theta-r-vp}{2}\widehat{K}(H+\theta+vp).
\end{split}
\end{equation}
The similar equations for the other two models are
\begin{equation}
H^2-v^2p^2+(2\theta-r)H+\theta(\theta-r)=-\frac{H+\theta-r+vp}{2}\widehat{K}(H+\theta-vp)-\frac{H+\theta-r-vp}{2}\widehat{K}(H+\theta+vp)
\label{eq:lapl-b}
\end{equation}
for Model B and
\begin{equation}
\left(H-\delta\right)^2-v^2p^2=-\frac{H-\delta+vp}{2}\widehat{K}(H-\delta-vp)-\frac{H-\delta-vp}{2}\widehat{K}(H-\delta+vp)
\label{eq:lapl-c}
\end{equation}
for Model C. In this more compact from the differences between the models are clearer, as the assumptions leading to varied parameter dependencies now take centre stage.\newline
\eqref{eq:laplaceform}-\eqref{eq:lapl-c} allow for immediate application for a multitude of kernels $\widehat{K}$. This is of particular note as one often does not know the equation for the kernel, but its Laplace transform is readily available. \eqref{eq:laplaceform} (or the corresponding equation for the other models) is employed to derive the front propagation velocity $u$ by requiring the functional $G(x(t),t)=0$ so that the total derivative $\frac{dG}{dt}=\frac{\partial G}{\partial t}+u\frac{\partial G}{\partial x}=0$. Consequently, $H(p)=up$ by using \eqref{eq: defns}. From Hamilton's equations we know that $\frac{\partial H}{\partial p}=\dot{q}=\frac{x}{t}$ where $x(t)=ut$ so we can set 
\begin{equation}
u=\frac{H}{p}=\frac{\partial H}{\partial p}.  
\label{eq:h-vel}
\end{equation}
This is equivalent to \cite{sergei2001frontvel}
\begin{equation}
u=\min_{p}\left( \frac{H}{p}\right).
\label{minvel}
\end{equation}
Having established the method by which the front propagation velocity is determined, we now proceed to apply \eqref{eq:h-vel} to our three models for different forms of the memory kernel $\widehat{K}$. 

\subsection{Forms of the Kernel}
\label{sec: kernel}
We now consider the effects of various kernels on the front propagation speed.\newline\newline
\textbf{Model A}\newline
\textit{Exponential running time PDF: }$\psi(\tau)=\lambda e^{-\lambda \tau}.$ In this case, the memory kernel $\widehat{K}=\lambda $ (see \eqref{eq: k}) leads to a Hamiltonian 
\begin{equation}
\left( H+\theta -r\right) ^{2}-v^{2}p^{2}=-\lambda \left( H+\theta
-r\right) ,
\end{equation}
which one can differentiate with respect to $p$ employing \eqref{eq:h-vel} to find 
\begin{equation}
H^{2}+H\left(\theta -r+\frac{\lambda}{2}\right)-v^{2}p^{2}=0.  
\label{eq:delta}
\end{equation}
From here we can find the derivative
\begin{equation}
\frac{\partial H}{\partial p}=\frac{2v^2p}{\sqrt{(\lambda/2+\theta-r)^2+4v^2p^2}},
\end{equation}
from which is trivially follows that if $\theta -r+\frac{\lambda}{2} =0$, the front propagation speed will be $u=\frac{\partial H}{\partial p}=v$. This corresponds to a system where the various rates affecting to the motion of the walker nullify each other, leaving the front to travel at the undisturbed individual velocity. This example is equivalent to that illustrated in \eqref{eq: fischer}.\newline
\textit{Gamma-distributed running time PDF: }$\psi(\tau)=\lambda^2\tau e^{-\lambda \tau}.$ The Laplace transform of this PDF is $\widehat{\psi}=\frac{\lambda^2}{(s+\lambda)^2}$. This leads to a memory kernel $\widehat{K}(s)=\frac{\lambda^{2}}{2\lambda +s}$ which now will depend on both $H,p$. Explicitly, the Hamiltonian is
\begin{equation}
\left(H+\theta-r\right)^2-v^2p^2=-\frac{\lambda^{2}\left(H+\theta-r+vp\right)}{2(2\lambda +H+\theta-vp)}-\frac{\lambda^{2}\left(H+\theta-r-vp\right)}{2(2\lambda +H+\theta+vp)}.
\label{eq: a-gamma}
\end{equation}
By the same method as before, one obtains the front velocity
\begin{equation}
\label{eq:result}
u=v\ \mathrm{provided}\ \lambda=\lambda_c=2(r-\theta)+\sqrt{2(r-\theta)^2+2r(r-\theta)}.
\end{equation}
This result is obtained by iteratively applying \eqref{eq:h-vel}. However, it does not preclude the existence of other minima. If $\lambda<2(r-\theta)+\sqrt{2(r-\theta)^2+2r(r-\theta)}$, the upper bound on the system is still $v$ for large $p$, but the front velocity is pushed more slowly to this limit. This is limited to values of $\lambda$ in the neighbourhood of \eqref{eq:result}. Alternatively, if $\lambda>2(r-\theta)+\sqrt{2(r-\theta)^2+2r(r-\theta)}$ the propagation velocity is less than the upper bound.

\begin{figure}
\centering
\begin{subfigure}{0.45\textwidth}
\centering
\includegraphics[scale=0.55]{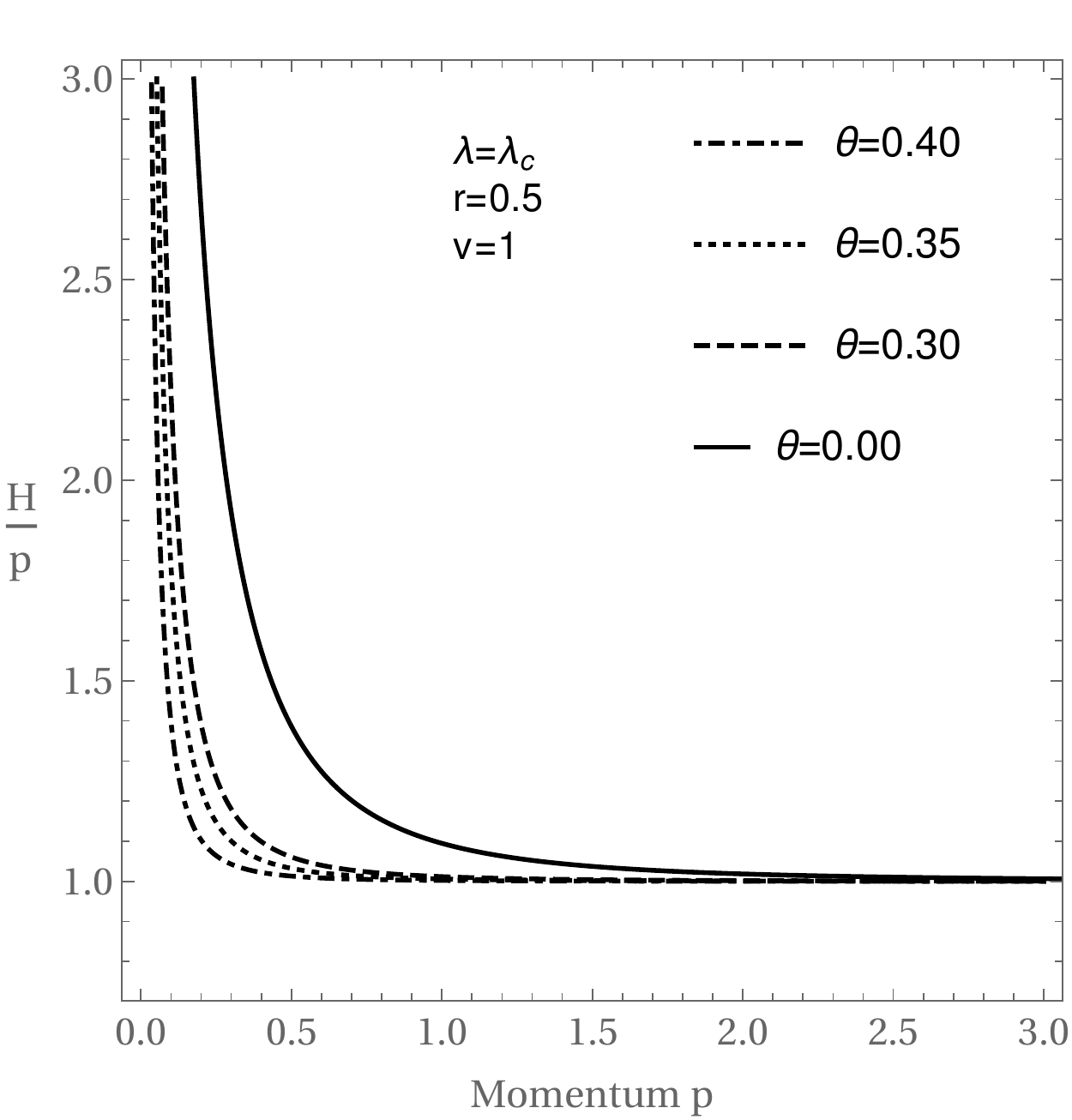}
\caption{$\widehat{K}(s)=\frac{\lambda^2}{2\lambda+s}$ for $\lambda=\lambda_c$.}
\end{subfigure}%
\begin{subfigure}{0.45\textwidth}
\centering
\includegraphics[scale=0.55]{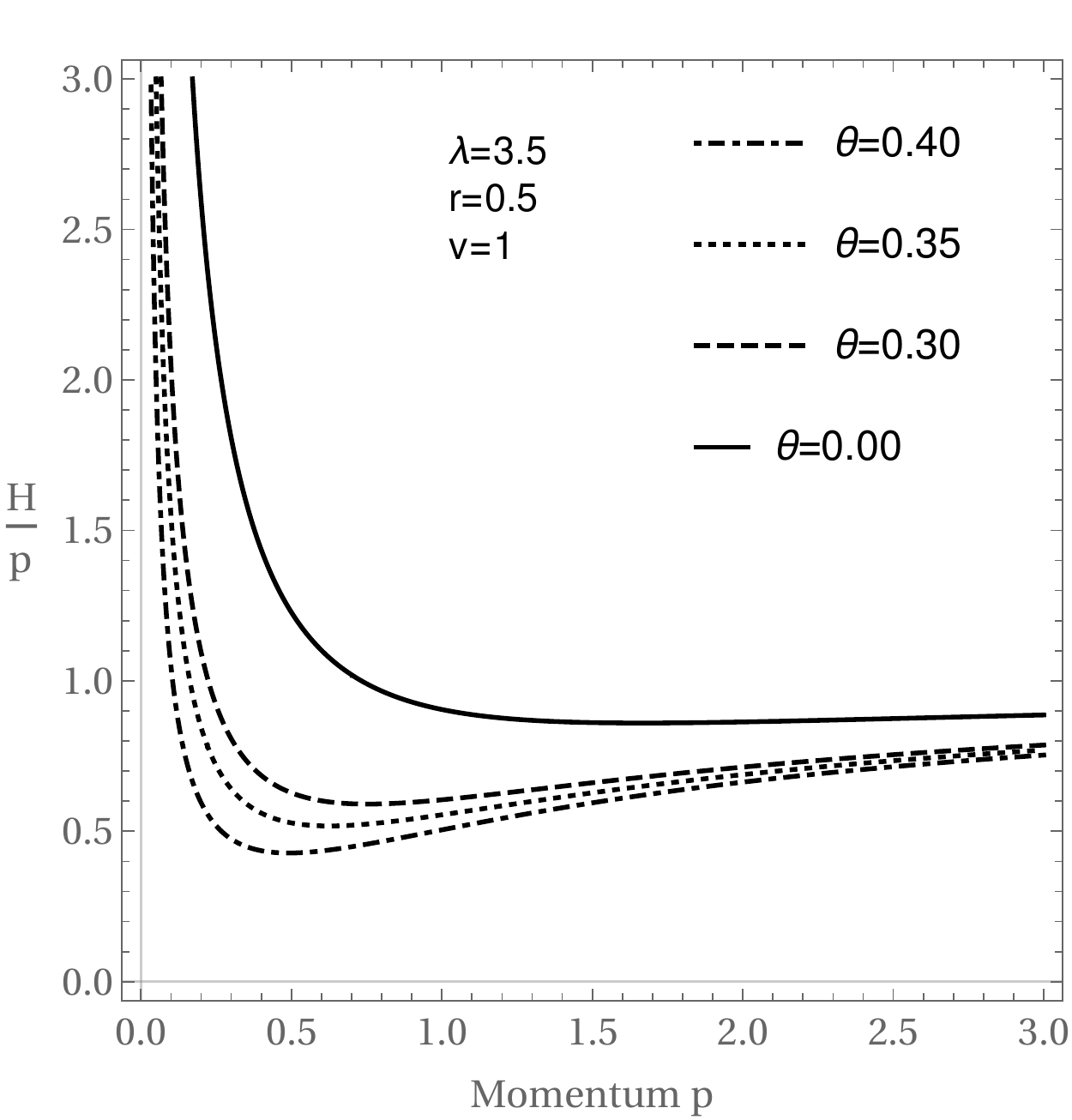}
\caption{$\widehat{K}(s)=\frac{\lambda^2}{2\lambda+s}$ for $\lambda=3.5$.}
\end{subfigure}
\caption{$\frac{H}{p}$ for a gamma-distributed kernel with various production-to-turning ratios. In the case of $\lambda$ obeying \eqref{eq:result}, the death rate has little effect on the propagation velocity and $\min_p(u)=v$. The curves tend to $u=v$ increasingly quickly as $\theta$ grows. However, in the case where the switching rate is larger all velocities have a minimum, even in the case of no death rate $\theta=0$. Now $\min_p(u)<v$ where the velocity is dependent on $\theta$. The details of this are defined by \eqref{eq:result}.}
\label{fig: Deathrates}
\end{figure}

Traditionally, the front velocity estimate from diffusion would be derived from a Hamiltonian of the form
\begin{equation}
H=Dp^2+r-\theta
\end{equation}
with consequent velocity of $u=2\sqrt{D(r-\theta)}$. Note that the Hamiltonian corresponds to \eqref{eq:eqn-f} for $\theta=0$.
The traditional velocity is an overestimate wrt. the result from \eqref{eq:result} when
\begin{equation}
r>\frac{17}{16}\theta,
\end{equation}
indicating that the growth term need not be much greater than the death rate before issues are encountered. In the case of \eqref{eq:delta}, the diffusion estimate is twice as large (see \eqref{eq:standard-v}) regardless of the values of the reaction parameters. Therefore, in cases where growth rates are relatively larger than death rates, a more careful estimate of the front velocity should be considered. We compare the velocities of this kernel for various $\lambda$ in Figure \ref{fig: Deathrates}, motivated by \eqref{minvel}.

\begin{figure}
\centering
\includegraphics[scale=.7]{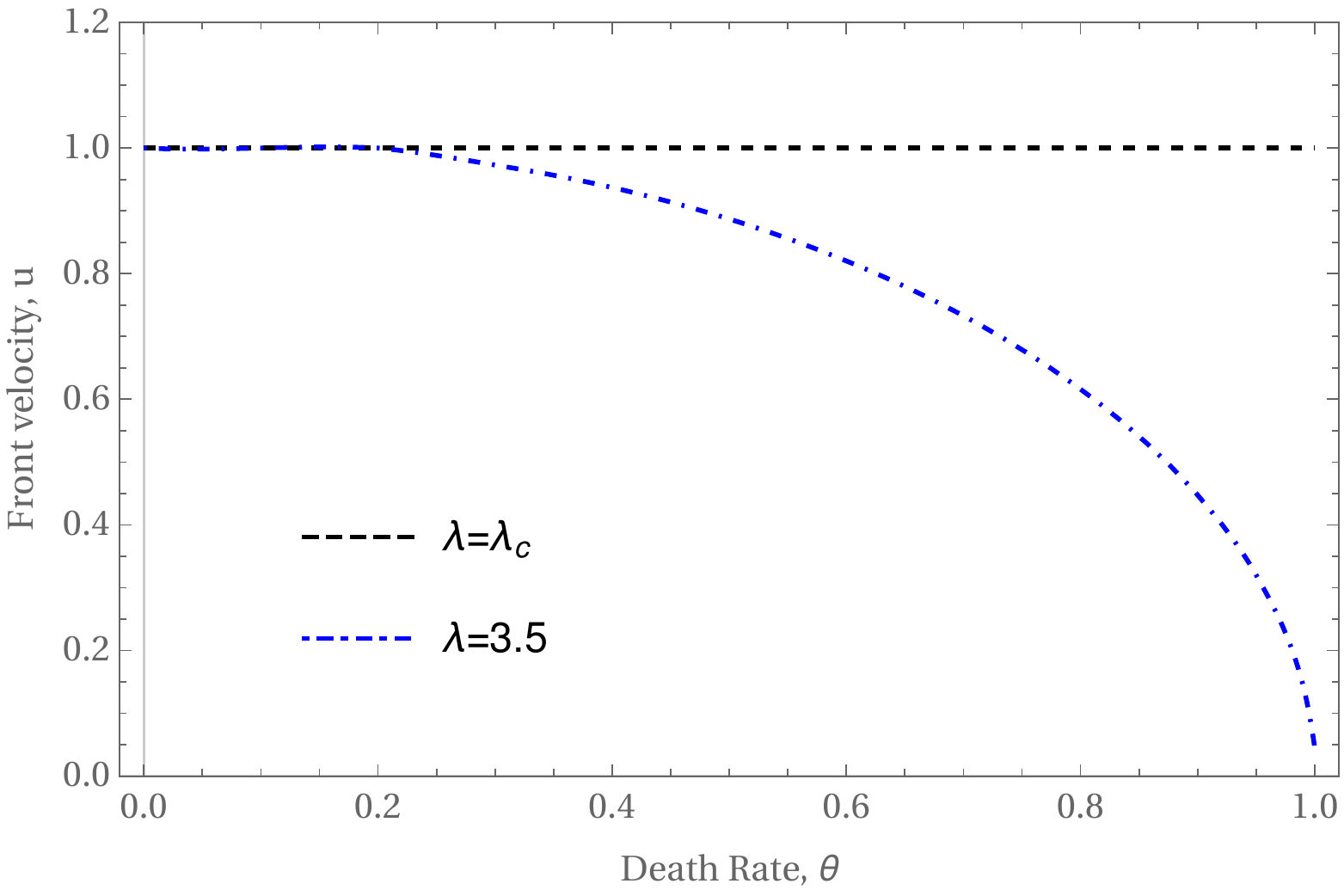}
\caption{Dependence of the front velocity (with a gamma-distributed memory kernel) on the death rate $\theta$ for $r=1, v=1$. In the case of $\lambda=\lambda_c=2(r-\theta)+\sqrt{2(r-\theta)^2+2r(r-\theta)}$, $\theta$ has no influence and we have $u=v$. However, in the case of $\lambda>\lambda_c$, $u=u(\theta)$ which decreases with the death rate.}
\label{fig: gamma-evol}
\end{figure}

Clearly there are conditions under which the minimum velocity will be the speed of the individuals walkers, and others where this is not the case. We illustrate the parameter dependence of these two cases in Figure \ref{fig: gamma-evol}.
\newline\newline
\textbf{Model B}\newline
\textit{Exponential running time PDF: }$\psi(\tau)=\lambda e^{-\lambda \tau}.$ In this case, the memory kernel $\widehat{K}=\lambda $ leads to a Hamiltonian
\begin{equation}
H^2-v^2p^2+(2\theta-r)H+\theta(\theta-r)=-(H+\theta-r)\lambda,
\end{equation}
from which is follows that $u=v$ provided $\lambda=r-2\theta>0$. This is different to the result obtained for Model A, where our assumptions were naturally satisfied by demanding $r>\theta$.\newline
\textit{Gamma-distributed running time PDF: }$\psi(\tau)=\lambda^2\tau e^{-\lambda \tau}$ with memory kernel $\widehat{K}(s)=\frac{\lambda^{2}}{2\lambda +s}$.
\begin{equation}
H^2-v^2p^2+(2\theta-r)H+\theta(\theta-r)=-\frac{\lambda^{2}\left(H+\theta-r+vp\right)}{2(2\lambda +H+\theta-vp)}-\frac{\lambda^{2}\left(H+\theta-r-vp\right)}{2(2\lambda +H+\theta+vp)}
\end{equation}
is again similar to \eqref{eq: a-gamma}, but lacks some of the symmetries in the expression of the parameters which previously allowed for simplification. By the same method as before, we obtain an expression for the front velocity using the parameter $q_p$ such that
\begin{equation}
\frac{\partial H}{\partial p}=\frac{4pv^2q_p}{\sqrt{16p^2v^2[3q_p+4\lambda(r+\lambda)]q_p+9(\theta+\lambda)^2[2q_p+4\lambda^2+r(\theta+5\lambda)]^2}},
\end{equation}
where $q_p=r\theta-2\theta^2-4\theta\lambda-3\lambda^2$. We note that this equation will still obey $\lim_{p\rightarrow\infty}\frac{\partial H}{\partial p}=v$, but that there is not a set of conditions which will ensure this for all $p$. Particularly for low values of $p$, it follows that $\frac{\partial H}{\partial p}<v$. Bearing in mind that this model corresponds to produced daughter particles moving in either direction, we understand the behaviour of $\frac{\partial H}{\partial p}<v$ as the result of a smaller contribution to the propagation front.\newline\newline
\textbf{Model C}\newline
\textit{Exponential running time PDF: }$\psi(\tau)=\lambda e^{-\lambda \tau}.$ Again, the memory kernel $\widehat{K}=\lambda $ yields
\begin{equation}
\left( H+\theta -r\right) ^{2}-v^{2}p^{2}=-\lambda \left( H+\theta-r\right),
\end{equation}
which is equivalent to the results obtained in Model A as expected from the form of $\widehat{K}$.\newline
\textit{Gamma-distributed running time PDF: }$\psi(\tau)=\lambda^2\tau e^{-\lambda \tau}$ with memory kernel $\widehat{K}(s)=\frac{\lambda^{2}}{2\lambda +s}$ gives the Hamiltonian
\begin{equation}
\left(H+\theta-r\right)^2-v^2p^2=-\frac{\lambda^{2}\left(H+\theta-r+vp\right)}{2(2\lambda +H+\theta-r-vp)}-\frac{\lambda^{2}\left(H+\theta-r-vp\right)}{2(2\lambda +H+\theta-r+vp)}.
\label{eq: b-gamma}
\end{equation}
By the same methods as employed in the other two models, we find that
\begin{equation}
u=v \text{ provided } \lambda=(r-\theta)(2+\sqrt{2}).
\end{equation}
In this case, we note that standard diffusion methods will always overestimate the propagation speed of the front, regardless of the values of $r, \theta$. The reason we are able to determine this is due to the symmetry in the parameters of \eqref{eq: b-gamma}. The resulting changes in the propagation speeds in these three models emphasize the large impact small variations can have on the final results. In particular, walkers produced with non-zero running time seem to have a large impact on the velocity estimates when compared to diffusion methods.

\vspace*{0.5cm}
\setcounter{equation}{0}
\section{Superdiffusive Motion}

We now consider the anomalous cases corresponding to power law distributions of the running times. These distributions are characterised by having heavy tails of the form
\begin{equation}
\psi(\tau)\sim\tau^{1-\mu},
\end{equation}
where $0<\mu<2$. We make a distinction between the strongly anomalous case occurring for $0<\mu<1$ and the anomalous L\'{e}vy walk occurring  for $1<\mu<2$. A key feature of anomalous distributions is the divergence of the first moment ($0<\mu<1$). For L\'{e}vy walks the first moment is finite, but the second moment diverges.\newline
If we assume the anomalous survival function takes the form of the Mittag-Leffler function
\begin{equation}
\Psi(\tau)=E_\alpha\left[-\left(\frac{\tau}{\tau_0}\right)^\mu\right],\qquad 0<\mu<1
\end{equation}
The Laplace transform of $\Psi(\tau)$ is
\begin{equation}
\widehat{\Psi}(s)=\frac{1}{s}\frac{(s\tau_0)^\mu}{1+(s\tau_0)^\mu},
\end{equation}
so that using \eqref{eq: k}, we obtain a memory kernel
\begin{equation}
\widehat{K}(s)=\frac{s^{1-\mu}}{\tau_0^\mu}.
\end{equation}
The Mittag-Leffler distribution is valid for all time, so that we need not worry about time scales. However, if the switching rate defined by \eqref{eq: gamma-def} is inversely proportional with the running time, 
\begin{equation}
\gamma(\tau)=\frac{\mu}{\tau_0+\tau}.
\end{equation}
This has running time
\begin{equation}
\Psi(\tau)=\left(\frac{\tau_0}{\tau+\tau_0}\right)^\mu,
\end{equation}
where $\mu$ and $\tau_0$ are constants \cite{newsergei}. Consequently, the kernel takes the form
\begin{equation}
\widehat{K}(s)=\frac{s^{1-\mu}}{\Gamma(1-\mu)\tau_0^\mu},
\label{eq: khat}
\end{equation}
which is only valid in the long time limit ($s\to0$). 
In the above case of \eqref{eq: khat} the behaviour does not apply in the case $\mu=1$ as a result of the gamma function $\Gamma(1-\mu)$.
To understand some of the properties of anomalous kernels, we first introduce an example based on the dynamics of Model A.\newline\newline
\textbf{Example}\newline
In this case, the governing equations are the same as our previous example in \eqref{eq: markov1}, \eqref{eq: markov2}
\begin{equation}
\frac{\partial\rho_+}{\partial t}+v\frac{\partial \rho_+}{\partial x}=-\frac{1}{2}(i_+ - i_-) + (f^+-\theta)\rho_+,
\end{equation}
\begin{equation}
\frac{\partial\rho_-}{\partial t}-v\frac{\partial \rho_-}{\partial x}=\frac{1}{2}(i_+ - i_-) + (f^+-\theta)\rho_-,
\end{equation}
but we cannot as easily transform back into $(x,t)$-space when defining the interaction terms. Consequently, we define it in inverse space as
\begin{equation}
\tilde{\imath}(k,s)=\tilde{\rho}(k,s)\frac{(s\mp ikv+\theta)^{1-\mu}}{\tau_0^\mu},\qquad 0<\mu<2.
\end{equation}
The two cases (strongly anomalous L\'{e}vy walk) are both included in this expression. Should we wish to, one can work in $(x,t)$-space by means of the fractional material derivative. For work in this area, see \cite{matderiv}. However, this is more involved than needed as our method explicitly deals with $\widehat{K}(s)$. Clearly for $\mu=1$ there are no memory effects and we return to a Markovian model. We now examine memory kernels for the two anomalous cases applied to Model A.

\subsection{Strong anomalous case}
\begin{figure}
\centering
\includegraphics[scale=0.7]{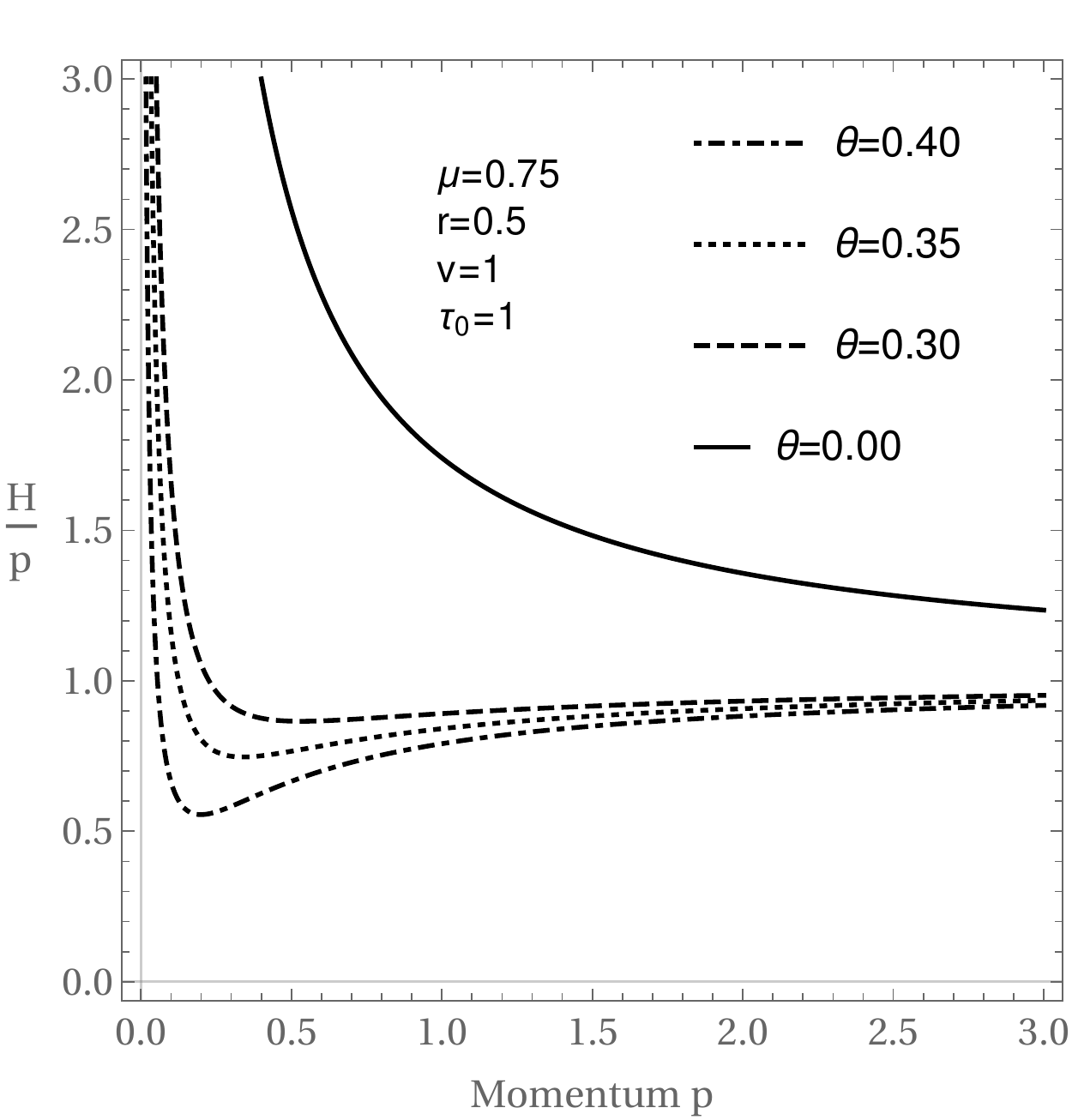}
\caption{$\frac{H}{p}$ for the strongly anomalous kernel $\widehat{K}(s)=\frac{s^{1-\mu}}{\tau_0^\mu}$ for $r=0.5,\ 0<\mu<1$. We compute these curves motivated by \eqref{minvel}, such that the minima correspond to the propagation velocity. The presence of the minima is independent of the values of $r$.}
\label{fig:anom75}
\end{figure}

\textit{Strong anomalous case: }$\psi(\tau) \sim \tau^{1-\mu },\ 0<\mu <1$. For the corresponding kernel $\widehat{K}=\frac{s^{1-\mu}}{\tau_0^\mu}$,  the Hamiltonian of Model A becomes
\begin{equation}
\left(H+\theta-r\right)^2-v^2p^2=-\left(H+\theta-r+vp\right)\frac{(H+\theta-vp)^{1-\mu}}{2\tau_0^\mu}-\left(H+\theta-r-vp\right)\frac{(H+\theta+vp)^{1-\mu}}{2\tau_0^\mu}.
\label{eq: mu-eq}
\end{equation}
The anomalous case is computed numerically and illustrated by Figure \ref{fig:anom75}. For ease of comparison with the gamma-distributed kernel, we employ the same values of $r,\theta$.

\begin{figure}
\centering
\includegraphics[scale=.7]{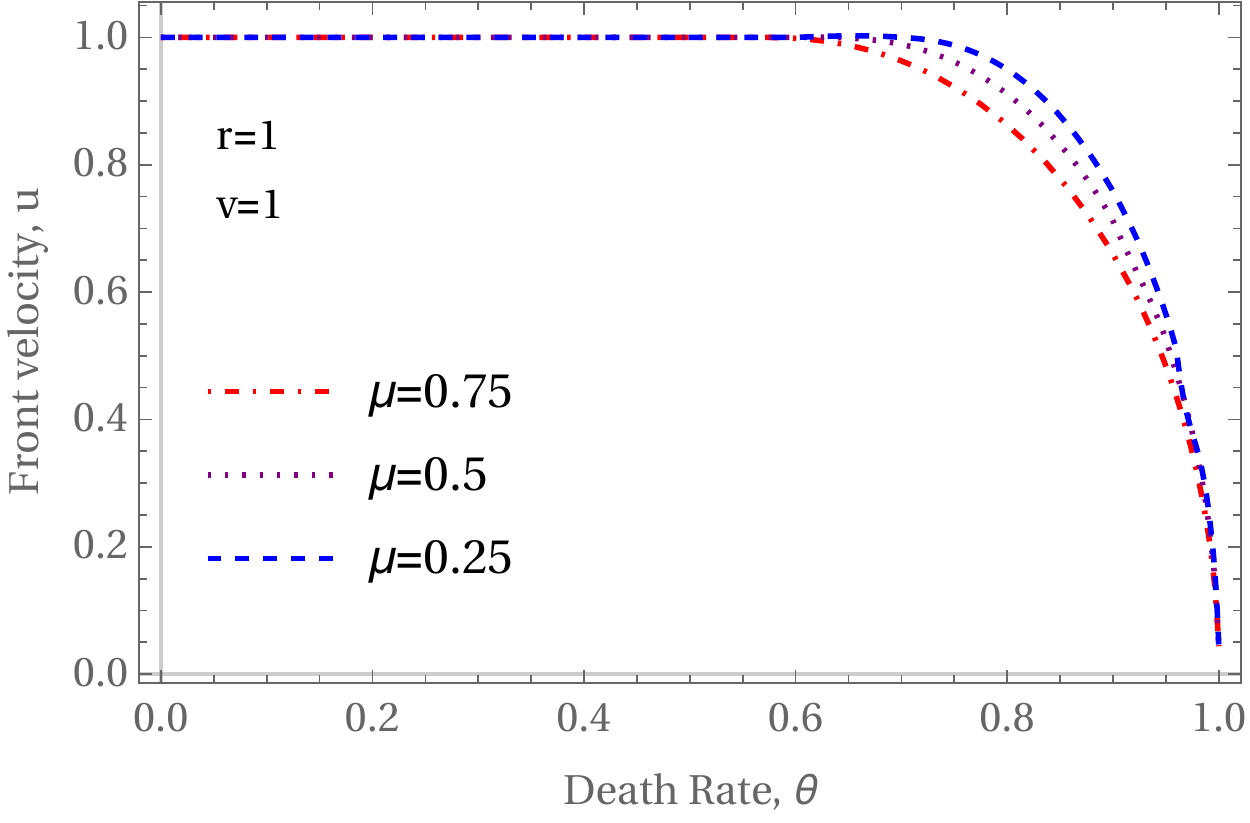}
\caption{Dependence of the front velocity (with a heavy-tailed memory kernel) on the death rate $\theta$. For small values of $\theta$ the front propagates with speed $v$. However, for $0.6<\theta<1$, the speed decreases towards 0, where the net production is null.
We note that the speed remains constant for longer as $\mu$ decreases, and that the range over which the speed changes is smaller than that for the gamma-distributed running time.}
\label{fig: g75-evol}
\end{figure}

We observe that the death rate not only modifies the propagation of the
system, but effectively `kills' the anomalous effects of the transport.
We can see from Figure \ref{fig:anom75} that the death rate has a large implication for the velocity. If the death rate is zero, the only minimum velocity is $v$.
For $\theta>0$ and a fixed birth rate, minima appear below $u=v$. This is a result of the coupling of the death rate with the turning kernel, and is unchanged by variations in the growth rate $r$ for the same model.\newline
This independence on $r$ is in contrast with the gamma-distributed kernel, where the behaviour of the propagation is dependent upon the relative values of $r,\lambda$. The dependence of these minima upon $\theta$ is examined in Figure \ref{fig: g75-evol} for several values of $0<\mu<1$.

\subsection{L\'{e}vy Walk}
\begin{figure}
\centering
\includegraphics[scale=0.7]{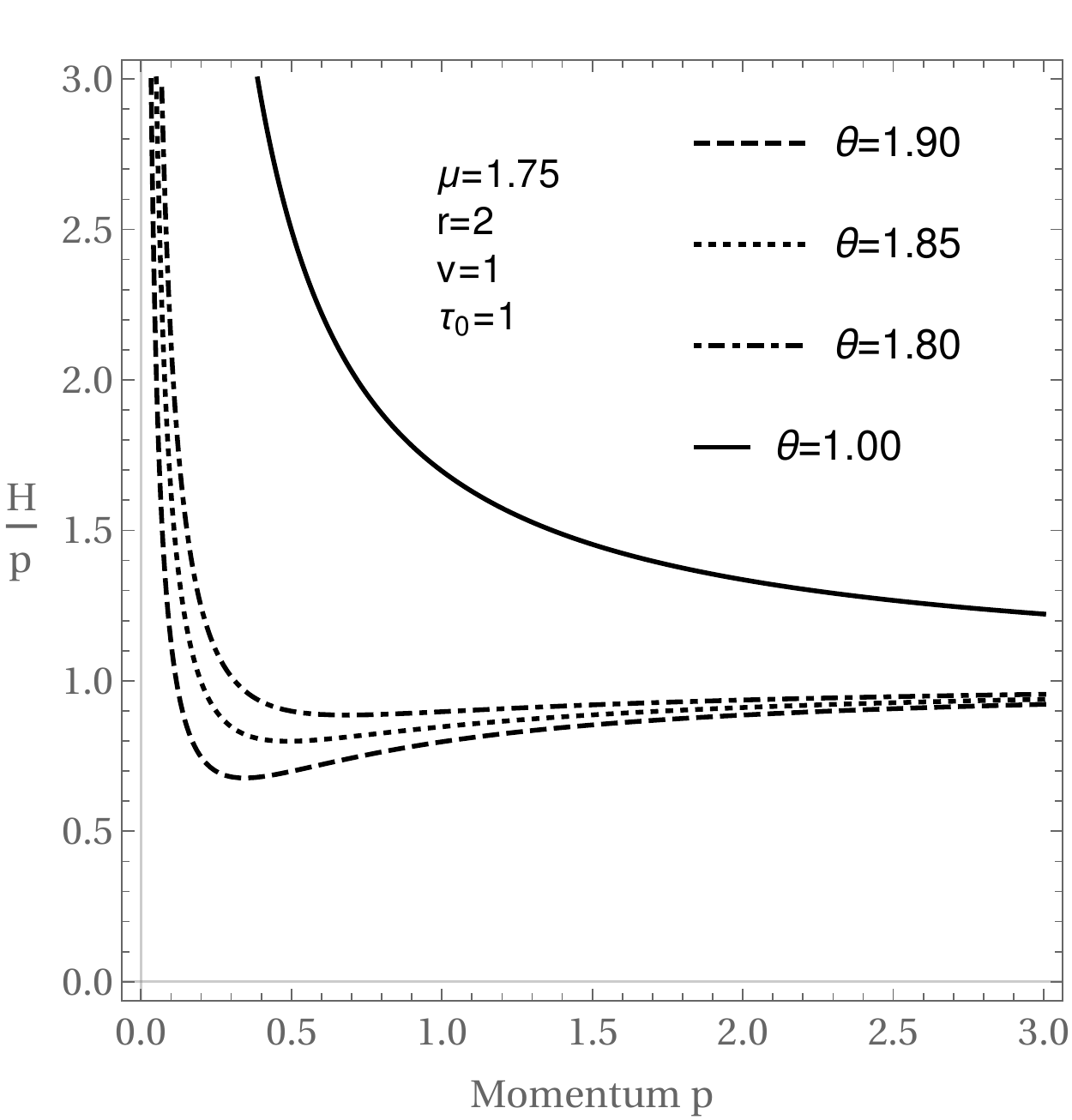}
\caption{$\frac{H}{p}$ for the anomalous kernel $\widehat{K}(s)=\frac{s^{1-\mu}}{\tau_0^\mu}$ for $r=2,\ 1<\mu<2$. We compute these curves in order to find the minima motivated by \eqref{minvel}. The presence of the minima is independent of the values of $r$. Note that our values of $r,\ \theta$ for this kernel are greater than those previously considered.}
\label{fig:anom175}
\end{figure}

\begin{figure}
\centering
\includegraphics[scale=.7]{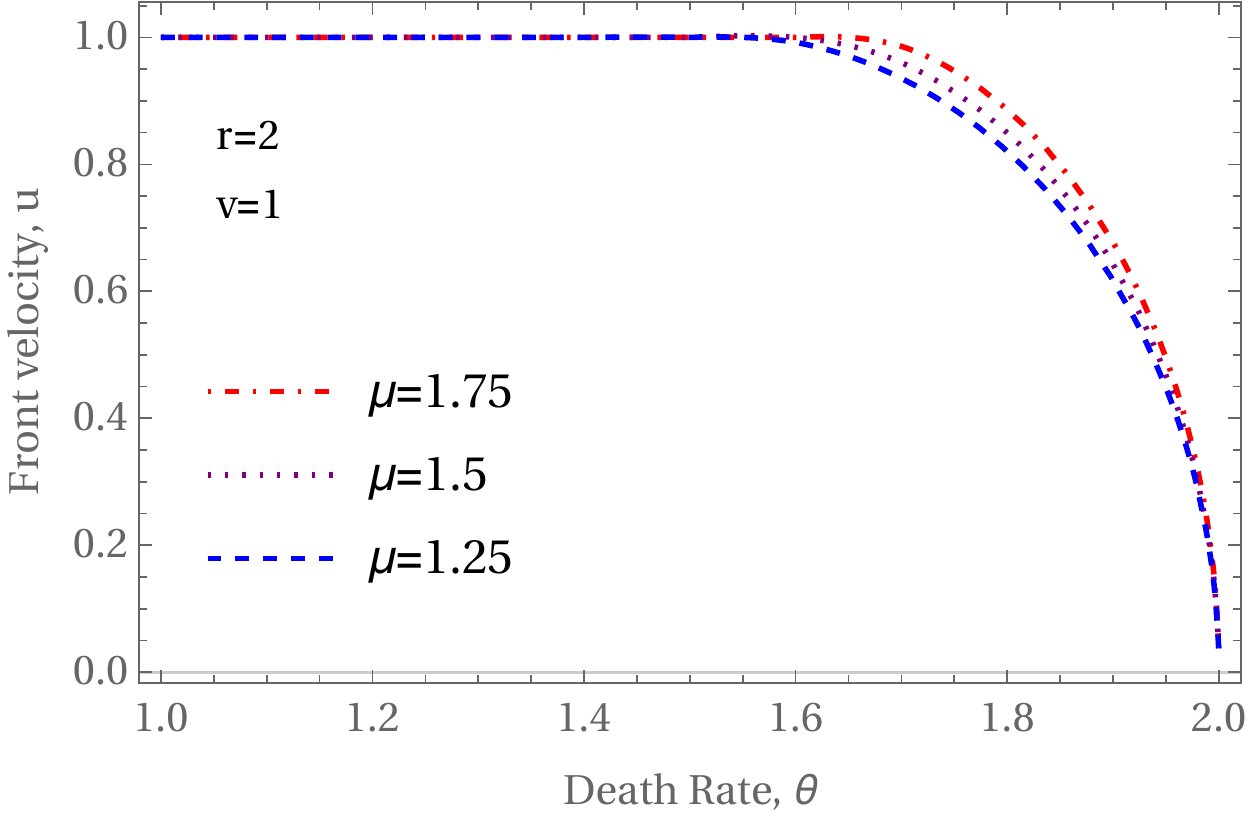}
\caption{Dependence of the front velocity on the death rate for $\theta>0$. The behaviour is similar to that observed in Figure \ref{fig: g75-evol}, though the domain of decrease in front velocity occurs over $1.7<\theta<2$. 
In this case, larger values of $\mu$ keep the speed at $u=v$ for a larger domain of $\theta$.}
\label{fig: anom175-evol}
\end{figure}

\textit{Anomalous case: }$\psi(\tau) \sim \tau^{1-\mu },\ 1<\mu <2$.\newline
The behaviour in this case has similar properties to that of the strongly anomalous case.
However, since $\mu>1$  we focus on $r,\ \theta>1$ in order to prevent very large terms dominating \eqref{eq: mu-eq}. From a physical perspective it is of interest to consider cases where birth and death rates are comparable, motivating an examination of larger $r,\theta$ in Figure \ref{fig:anom175}. 
\newline
For a fixed birth rate, we again observe minima below $v$, which we examine in greater detail in Figure \ref{fig: anom175-evol}. In this case, for comparable values of $r, \theta$ to the strongly anomalous kernel, the minima occur over a smaller range of $\frac{H}{p}$. As the minimum must eventually reach zero, this indicates the change in front velocity should be `faster' than in the previous case. 

We note that both the anomalous cases have displayed a `faster' decrease in the velocity compared to the gamma-distributed kernel (the decrease happens over a shorter range of $\theta$). The heavy tails of the distribution disappear when accounting for a non-zero death rate. This corresponds to the death rate tempering the heavy tails of the power law distributions and effectively `killing' the very long running times. 

Our results clearly indicate that the death term has a significant impact on the front propagation speeds. In particular, the smaller velocity $u<v$ was observed to occur independently of the values of $r$, provided the assumptions laid out in Section 2 were maintained.
If we na\"{i}vely were to consider death by simply reducing the magnitude of the birth term, these key effects would be lost.

\vspace*{0.5cm}
\setcounter{equation}{0}
\section{Conclusion}
We have formulated three different models for L\'{e}vy walkers with general running time distributions with birth and death processes. We have further derived the corresponding integro-differential equations for the mean density for all three cases. In order to obtain the travelling wave solution we applied a hyperbolic transformation from the onset without further diffusion approximations. We have determined the Hamilton-Jacobi equations for these models and consequently equations for the fronts of the travelling waves. 

Where possible, we found the conditions under which the parameters of the models led to a front propagation velocity equal to the speed of the walkers. These conditions depend on the equation for the memory kernel, and were shown to vary significantly depending on the assumptions employed in the three models. Consequently, we found conditions under which standard diffusion methods produce an overestimate of the propagation velocity.

In the anomalous case involving power-law running time distributions we found that the death rate has a significant impact on the front propagation, making it less than the maximum speed of the walkers. We interpret this observation as the death rate tempering the anomalous effects in the transport by effectively `killing' the very long running times. 
In the strongly anomalous case of $0<\mu<1$, the front velocity was shown to decrease as $\theta\to r$. For smaller values of $\mu$, the front velocity remained at $u=v$ for a larger domain on $\theta$.
In the case of $1<\mu<2$, the front velocity was also observed to change as $\theta\to r$, but $u=v$ was observed for a larger domain of $\theta$ as $\mu$ increased. This suggests a constant front propagation $u=v$ for $\mu=0,\ 2$.\newline
Is is important to note that this behaviour would not have been observed had we na\"{i}vely worked with a net birth term rather than accounting for the effects separately.
The tempering of the anomalous behaviour effected by the death term is in contrast with the slowing down one would expect from simpler models using an approach of the form \eqref{eq:standard-v}.\newline
Our models have introduced birth and death processes directly into the transport process. For biologically motivated models, it may be of interest to consider these processes as occurring during a resting state. Our method can easily be extended to account for this by introducing a distribution of resting times in addition to the running time $\tau$.

\renewcommand{\theequation}{\Alph{section}.\arabic{equation}}
\setcounter{equation}{0}
\setcounter{section}{0}
\section{Appendix A}
In this appendix we derive the non-Markovian master equations for the densities $\rho_\pm$ and determine the structure of the exchange terms which describe the changes in direction of the individuals. If we simply denote the number of walkers moving in either direction as $n_\pm(x,t,\tau)$, we can write the equations of motion
\begin{equation}
\label{eq:nb}
\frac{\partial n_+}{\partial t}+\frac{\partial n_+}{\partial\tau}+v\frac{\partial n_+}{\partial x}=-\gamma(\tau)n_+ -\theta n_+ +f^+n_+,
\end{equation}
\begin{equation}
\label{eq:n2}
\frac{\partial n_-}{\partial t}+\frac{\partial n_-}{\partial\tau}-v\frac{\partial n_-}{\partial x}=-\gamma(\tau)n_- -\theta n_- + f^+n_-,
\end{equation}
where $\gamma(\tau)$ is the switching rate dependent on the running time so that our equations depend on $\tau$. As our running time initial conditions for the birth term $f^+$ result in its exact form having no influence on our equations, for simplicity we do not here concern ourselves with its density dependence. We want to eliminate the variable $\tau$ and do so by defining the density
\begin{equation}
\rho_\pm(x,t)=\int_0^tn_\pm(x,t,\tau)d\tau.
\end{equation}
We will need the conditions
\begin{equation}
n_\pm(x,0,\tau)=\frac{1}{2}(\rho_+^0(x)+\rho_-^0(x))\delta(\tau),\quad n_\pm(x,t,0)=\frac{1}{2}\int_0^t\gamma(\tau)(n_-(x,t,\tau)+n_+(x,t,\tau))d\tau
\end{equation}
where $\rho_\pm^0(x)$ are the initial distributions.
Using the method of characteristics, $n_\pm(x,t,\tau)=n_\pm(x\mp v\tau,t-\tau,0)e^{-\theta\tau}e^{-\int_0^\tau\gamma(u)du}$. Choosing to write
\begin{equation}
\Psi(\tau)=e^{-\int_0^\tau\gamma(u)du}=-\frac{\Psi'(\tau)}{\gamma(\tau)},
\end{equation}
and noting that by definition $i_\pm=\int_0^t\gamma(\tau)n_\pm(x,t,\tau)d\tau$, so that $n_\pm(x,t,0)=\frac{1}{2}(i_++i_-)$. By substitution, this yields
\begin{equation}
\rho_\pm(x,t)=\frac{1}{2}\int_0^t\left[i_++i_-\right](x\mp v\tau,t-\tau)\Psi(\tau)e^{-\theta\tau}d\tau+\frac{1}{2}\left[\rho_+^0+\rho_-^0\right](x\mp vt)\Psi(t),
\end{equation}
\begin{equation}
i_\pm(x,t)=\frac{1}{2}\int_0^t\left[i_++i_-\right](x\mp v\tau,t-\tau)\psi(\tau)e^{-\theta\tau}d\tau+\frac{1}{2}\left[\rho_+^0+\rho_-^0\right](x\mp vt)\psi(t)\\
\end{equation}
where on the last line we have used the result $\Psi'(\tau)=-\psi(\tau)$ to eliminate $\gamma(\tau)$.

Applying the Fourier-Laplace transform to these equations, we get
\begin{equation}
\tilde{\imath}_\pm(k,s)=\frac{1}{2}[\tilde{\imath}_+(k,s)+\tilde{\imath}_-(k,s)+\rho_+^0(k)+\rho_-^0(k)]\hat{\psi}(s\mp ikv+\theta),
\end{equation}
\begin{equation}
\tilde{\rho}_\pm(k,s)=\frac{1}{2}[\tilde{\imath}_+(k,s)+\tilde{\imath}_-(k,s)+\rho_+^0(k)+\rho_-^0(k)]\hat{\Psi}(s\mp ikv+\theta),
\end{equation}
from which it follows
\begin{equation}
\tilde{\imath}_\pm(k,s)=\frac{\hat{\psi}(s\mp ikv+\theta)}{\hat{\Psi}(s\mp ikv+\theta)}\tilde{\rho}(k,s)=\hat{K}(s\mp ikv+\theta)\tilde{\rho}(k,s)
\end{equation}
from \eqref{eq: k}. Taking the inverse Laplace transform, we obtain \eqref{eq:interaction} using the shift theorem. We note that under the assumption that new individuals are produced with zero running time, the form of the production term is irrelevant and so the derivation also applies for sufficiently well-behaved $f^+=f^+(\rho)$.

\setcounter{equation}{0}
\section{Appendix B}
This appendix finds the full integro-differential equation of the total density of walkers, using methods previously considered in \cite{HillenDL2000}. From \eqref{eq:rho1-b}, \eqref{eq:rho2-b} we add and subtract the terms,
\begin{equation}
\label{eq:dJx}
\frac{\partial(\rho_++\rho_-)}{\partial t}+v\frac{\partial(\rho_+-\rho_-)}{\partial x}=\frac{\partial\rho}{\partial t}+\frac{\partial J}{\partial x}=(f^+-\theta)(\rho_++\rho_-)=(f^+-\theta)\rho
\end{equation}
\begin{equation}
\begin{split}
\frac{\partial(\rho_+-\rho_-)}{\partial t}+v\frac{\partial(\rho_++\rho_-)}{\partial x}=\frac{1}{v}\frac{\partial J}{\partial t}+v\frac{\partial \rho}{\partial x}=-&(i_+-i_-)+(f^+-\theta)(\rho_+-\rho_-)\\
=-&(i_+-i_-)+(f^+-\theta)\frac{J}{v},
\end{split}
\end{equation}
where the second forms arise from the definitions of $J=v\rho_+-v\rho_-,\ \rho=\rho_++\rho_-$. By cross-differentiation wrt. space and time, 
\begin{equation}
\frac{\partial^2\rho}{\partial t^2}+\frac{\partial^2 J}{\partial x\partial t}=(f^+-\theta)\frac{\partial\rho}{\partial t}
\end{equation}
\begin{equation}
\begin{split}
v^2\frac{\partial^2\rho}{\partial x^2}+\frac{\partial^2 J}{\partial x\partial t}=-&v\frac{\partial}{\partial x}(i_+-i_-)+(f^+-\theta)\frac{\partial J}{\partial x}\\
=-&v\frac{\partial}{\partial x}(i_+-i_-)+(f^+-\theta)^2\rho-(f^+-\theta)\frac{\partial\rho}{\partial t}
\end{split}
\label{eq: working}
\end{equation}
where we have used \eqref{eq:dJx} in the last step. Hence we can eliminate the remaining flux term $\frac{\partial^2 J}{\partial x\partial t}$ to obtain \eqref{eq:diffeq}. Using \eqref{eq:interaction}, \eqref{eq: working} and \eqref{eq: rhopm}, we can write
\begin{equation}
\begin{split}
-v\frac{\partial}{\partial x}(i_+-i_-)=-&\frac{1}{2}\int_0^tK(t-\tau)\left[v\frac{\partial\rho}{\partial x}+\frac{\partial J}{\partial x}\right](x-v(t-\tau),\tau)e^{-\theta(t-\tau)}d\tau\\
+&\frac{1}{2}\int_0^tK(t-\tau)\left[v\frac{\partial\rho}{\partial x}-\frac{\partial J}{\partial x}\right](x+v(t-\tau),\tau)e^{-\theta(t-\tau)}d\tau,\\
=+\frac{1}{2}\int_0^tK(t-\tau)&e^{-\theta(t-\tau)}\left(\frac{\partial}{\partial t}+\theta-f^+-v\frac{\partial}{\partial x}\right)\rho(x-v(t-\tau),\tau)d\tau\\
+\frac{1}{2}\int_0^tK(t-\tau)&e^{-\theta(t-\tau)}\left(\frac{\partial}{\partial t}+\theta-f^++v\frac{\partial}{\partial x}\right)\rho(x+v(t-\tau),\tau)d\tau,
\end{split}
\end{equation}
so that \eqref{eq:start} is no longer flux dependent.

\setcounter{equation}{0}
\section{Appendix C}
This appendix discusses an alternative method by which the Hamilton-Jacobi equation of our two-state systems may be found. By the same exponential transformations as before we write
\begin{equation}
\rho_+(x,t)=Ae^{-\frac{G}{\epsilon}}, \qquad \rho_-(x,t)=Be^{-\frac{G}{\epsilon}},
\end{equation}
which we substitute directly into \eqref{eq:rho1-b}, \eqref{eq:rho2-b}. Together with a hyperbolic transformation, this yields the governing equations
\begin{equation}
-\frac{\partial G}{\partial t}\rho_+^\epsilon-v\frac{\partial G}{\partial x}\rho_+^\epsilon=-\theta\rho_+^\epsilon+f^+(\rho^\epsilon)\rho_+^\epsilon-\frac{1}{2}(i_+^\epsilon-i_-^\epsilon),
\end{equation}
\begin{equation}
-\frac{\partial G}{\partial t}\rho_-^\epsilon+v\frac{\partial G}{\partial x}\rho_-^\epsilon=-\theta\rho_-^\epsilon+f^+(\rho^\epsilon)\rho_-^\epsilon+\frac{1}{2}(i_+^\epsilon-i_-^\epsilon),
\end{equation}
with the interaction terms
\begin{equation}
i^\epsilon_\pm=\int_0^{t/\epsilon}K(\tau)e^{-\theta\tau}\rho_\pm^\epsilon\left(\frac{x}{\epsilon}\mp v\tau,\frac{t}{\epsilon}-\tau\right)d\tau.
\end{equation}
By expansion to first order in $\epsilon$, these equations take the form
\begin{equation}
-\frac{\partial G}{\partial t}\rho_+^\epsilon-v\frac{\partial G}{\partial x}\rho_+^\epsilon=-\theta\rho_+^\epsilon+r\rho_+^\epsilon-\frac{\rho_+^\epsilon}{2}\widehat{K}\left(-\frac{\partial G}{\partial t}-v\frac{\partial G}{\partial x}+\theta\right)+\frac{\rho_-^\epsilon}{2}\widehat{K}\left(-\frac{\partial G}{\partial t}+v\frac{\partial G}{\partial x}+\theta\right),
\end{equation}
\begin{equation}
-\frac{\partial G}{\partial t}\rho_-^\epsilon+v\frac{\partial G}{\partial x}\rho_-^\epsilon=-\theta\rho_-^\epsilon+r\rho_-^\epsilon+\frac{\rho_+^\epsilon}{2}\widehat{K}\left(-\frac{\partial G}{\partial t}-v\frac{\partial G}{\partial x}+\theta\right)-\frac{\rho_-^\epsilon}{2}\widehat{K}\left(-\frac{\partial G}{\partial t}+v\frac{\partial G}{\partial x}+\theta\right).
\end{equation}
These can be expressed in matrix form
\begin{equation}
\begin{split}
&M=\begin{pmatrix}
-\frac{\partial G}{\partial t}-v\frac{\partial G}{\partial x}+\theta-r+\frac{1}{2}\widehat{K}\left(-\frac{\partial G}{\partial t}-v\frac{\partial G}{\partial x}+\theta\right) & -\frac{1}{2}\widehat{K}\left(-\frac{\partial G}{\partial t}+v\frac{\partial G}{\partial x}+\theta\right)\\
-\frac{1}{2}\widehat{K}\left(-\frac{\partial G}{\partial t}-v\frac{\partial G}{\partial x}+\theta\right) & -\frac{\partial G}{\partial t}+v\frac{\partial G}{\partial x}+\theta-r+\frac{1}{2}\widehat{K}\left(-\frac{\partial G}{\partial t}+v\frac{\partial G}{\partial x}+\theta\right)
\end{pmatrix},\\
&M
\begin{pmatrix}
A\\
B
\end{pmatrix}=
\begin{pmatrix}
0\\
0
\end{pmatrix}
\end{split}
\end{equation}
where non-trivial solutions arise from the requirements that the determinant of the matrix be zero. By direct evaluation,
\begin{equation}
\begin{split}
&\left(-\frac{\partial G}{\partial t}-v\frac{\partial G}{\partial x}+\theta-r+\frac{1}{2}\widehat{K}\left(-\frac{\partial G}{\partial t}-v\frac{\partial G}{\partial x}+\theta\right)\right)\times\\
&\left(-\frac{\partial G}{\partial t}+v\frac{\partial G}{\partial x}+\theta-r+\frac{1}{2}\widehat{K}\left(-\frac{\partial G}{\partial t}+v\frac{\partial G}{\partial x}+\theta\right)\right)-\\
&-\frac{1}{4}\widehat{K}\left(-\frac{\partial G}{\partial t}+v\frac{\partial G}{\partial x}+\theta\right)\widehat{K}\left(-\frac{\partial G}{\partial t}-v\frac{\partial G}{\partial x}+\theta\right)=0.
\end{split}
\end{equation}
This simplifies to \eqref{eq:elimit-b}.

\section*{Acknowledgements}
S.F. acknowledges the support of the EPSRC Grant No. EP/J019526/1 "Anomalous reaction-transport equations".\newline
V.M. acknowledges the support of FIS2012-32334 by the "Ministerio de Econom\'{i}a y Competitividad" (Spain).

\printbibliography

\end{document}